\documentclass[11pt]{article}
\usepackage{epsfig}
\usepackage{graphicx,psfrag}
\usepackage{amssymb, amsmath}
\usepackage{amsthm,bm}
\usepackage{verbatim}
\usepackage{asa1}
\usepackage{color}
\usepackage{multirow}
\usepackage{colortbl}
\usepackage{float}
\begin{document}

\begin{center}
{\Large \bf Global multivariate point pattern models for rain type occurrence }

{\large  Mikyoung Jun}\footnote{Mikyoung Jun is an Associate Professor, Department of Statistics, Texas A\&M University, 3143 TAMU, College Station, TX 77843-3143 (E-mail: mjun@stat.tamu.edu). 
Mikyoung Jun acknowledges the support by NSF DMS-1208421 and DMS-1613003, and helpful discussion with Yongtao Guan on point process models. 
},
{\large  Courtney Schumacher}\footnote{Courtney Schumacher is a Professor, Department of Atmospheric Sciences, Texas A\&M University, and acknowledges support from NASA PMM grant NNX16AE34G.}, {\large  and R. Saravanan}\footnote{R. Saravanan is a Professor, Department of Atmospheric Sciences, Texas A\&M University, and acknowledges support from NSF AGS-1347808. \\
The authors also acknowledge an Interdisciplinary Seed Grant in Big Data program from Texas A\&M University. TRMM satellite data were provided by the NASA/Goddard Space Flight
Center and PPS. MERRA was developed by the Global Modeling and
Assimilation Office and supported by the NASA Modeling, Analysis and
Prediction Program. Source files for the TRMM PR and MERRA-2 data can be
acquired from the Goddard Earth Science Data Information Services Center
(GES DISC) (https://disc.gsfc.nasa.gov/).\\
Aaron Funk processed the TRMM PR and MERRA-2 data onto coincident
temporal and spatial grids. Junho Yang calculated EOFs used in the analysis.} 

\end{center}

%\begin{abstract}
{\bf Abstract}: We seek statistical methods to study the occurrence of multiple rain types observed by satellite on a global scale. The main scientific interests are to relate rainfall occurrence with various atmospheric state variables and to study the dependence between the occurrences of multiple types of rainfall (e.g. short-lived and intense versus long-lived and weak; the heights of the rain clouds are also considered). Commonly in point process model literature, the spatial domain is assumed to be a small, and thus planar domain. We consider the log-Gaussian Cox Process (LGCP) models on the surface of a sphere and take advantage of cross-covariance models for spatial processes on a global scale to model the stochastic intensity function of the LGCP models. We present analysis results for rainfall observations from the TRMM satellite and atmospheric state variables from MERRA-2 reanalysis data over the tropical Eastern and Western Pacific Ocean, as well as over the entire tropical and subtropical ocean regions. Statistical inference is done through Monte Carlo likelihood approximation for LGCP models. We employ covariance approximation to deal with massive data. 

\vspace{.5cm}

{ KEY WORDS}: Global spatial data; Log-Gaussian Cox process; Point process models; Rainfall occurrence; TRMM precipitation radar
%\end{abstract}

\newpage

\section{Introduction}
 
%(make sure all the results are updated for JASA)
%(comments from the presentation -- point vs grid)

%Spatial or spatio-temporal point pattern data commonly occur in environmental and health applications. Earthquake occurrences have been modeled extensively through point process models such as Hawkes process (e.g. \citenp{fox_et_al16}). Both wildfire and human-induced fire ignition data are modeled using various point process models (e.g. \citenp{genton_et_al06}, \citenp{ } ).  and  wild fire, crime, disease, invasive species, 911 call.

%Despite numerous scientific applications that can benefit from statistical analysis of point patterns in spatial and spatio-temporal context, literature on statistical methods and models for dealing with those point patterns (especially in multivariate context, or on a global scale) has been scarce. This is especially true when compared to active research activities regarding geostatistical xxx.

%The motivating scientific problem for this paper is tri-variate rainfall occurrences data on a global scale. (background on rainfalls)

Despite tremendous efforts by researchers to understand the global atmospheric circulation and climate, state-of-the-art climate models, that is General Circulation Models (GCMs), still exhibit pervasive biases. For example, climate models used for understanding the human influence on climate change, namely the Coupled Model Intercomparison Project phase 3 and 5 models (CMIP3 and CMIP5, respectively), show only a slight improvement in terms of their representation of rainfall (\citenp{flato_et_al13}).
Accurate understanding of rainfall distribution over space and time is crucial, as it is not just a matter of local rainfall but entails the forcing of atmospheric circulations around the globe (\citenp{hartmann_et_al84,schumacher_et_al04}) and the sensitivity to anthropogenic climate change (\citenp{sherwood_et_al14}). Poor rainfall representation in models also degrades the simulation of tropical phenomena such as the Madden-Julian Oscillation (MJO) and El Ni\~{n}o that contribute to atmospheric predictability (\citenp{hung_et_al13,zhu_et_al17}).

Over the last 19 years, high quality measurements of rainfall over the tropics and extratropics have become available via NASA's Tropical Rainfall Measurement Mission (TRMM; \citenp{kummerow_et_al98}) and the Global Precipitation Measurement (GPM; \citenp{hou_et_al14}) mission satellites. These high quality data sets help provide us better understanding of rainfall characteristics around the globe, which can help improve climate model simulations of rainfall. The radars onboard the TRMM and GPM satellites provide rainfall occurrence and amount for three different types of rain, namely {\it stratiform}, {\it deep convective}, and {\it shallow convective}. Each of these rain types have different properties in terms of intensity, duration, and height in the atmosphere and are further described in Section~\ref{sec:rainfall}.

Little work has been done in developing flexible statistical models and methods to understand how rainfall happens, let alone how different rain types are characterized and interact with each other. Many statistical studies regarding rainfall focus on relatively small regional domains (e.g., \citenp{seo99,frei_schar01,cowpertwait_et_al07,sun_stein15}). % analyzed 12 rain gauges data over parts of Virginia, Maryland, and North Carolina.
 Also most of these studies focus on the rainfall data itself, with less focus on understanding how rainfall is related to atmospheric state variables such as temperature and humidity. These state variables have strong physical connections to rainfall amounts and rain types (e.g., \citenp{johnson_et_al99,bretherton_et_al04,ahmed_schumacher15}), and statistical modeling of these connections can shed light on the processes that control rainfall occurrence and strength. This underscores the need to develop flexible statistical models to understand not only how each rain type occurs but also the joint distributional structure for these three different types of rainfall. %In this study, we are interested in the spatial location of rainfall occurrence for the three rain types and how they are related to various atmospheric state variables.
%We plan to develop spatial point process models for multivariate point patterns on a global scale. %These models are nonstationary in the sense that the intensity function of point processes for rainfall occurrences, tends to vary over space through atmospheric state variables that vary over space.

Statistical methods for point processes are concerned with arrangements (or patterns) of points in a random set (temporal, spatial or spatio-temporal domain). There are numerous types of data that come as point patterns in physical, environmental, and biological applications. 
Statistical methods for spatial (or spatio-temporal) point patterns have been developed for various aspects of the analyses, such as stochastic models and methods (\citenp{moller_et_al98,schlather_et_al04}), model fitting and inference (\citenp{diggle85,guan06,waagepeterson_guan09}), and goodness-of-fit methods for the statistical models (\citenp{guan08}). Various point process models have been used in a wide variety of applications (\citenp{schoenberg03,diggle_et_al05,peng_et_al05,zammit_mangion_et_al12}). \cite{diggle} as well as \cite{moller_waagepetersen04} provide nice overviews of the field with further references. 

In statistical modeling of spatial point patterns, a Poisson process often serves as a building block for more complex models. When the intensity function of a Poisson process is constant over the spatial domain we call it a {\it homogeneous} Poisson process, and when it varies over space we call it an {\it inhomogeneous} Poisson process. Spatial point pattern data in most real applications are not suitable for being modeled with homogeneous Poisson process models due to the models' obvious limitations of spatially constant intensity functions. 

One prominent approach for dealing with inhomogeneous spatial point patterns is through the so-called {\it Cox process} (or ``doubly stochastic'' process, \citenp{cox55}).
A spatial Cox process in a planar domain, $D\subset \mathbb{R}^2$, is defined via the following two properties:
\begin{enumerate}
\item $\{\Lambda(\mathbf{x}):\mathbf{x} \in D \}$ is a non-negative-valued stochastic process 
\item conditional on $\{\Lambda(\mathbf{x}) = \lambda(\mathbf{x}): \mathbf{x} \in D \}$, the event from an inhomogeneous Poisson process with intensity function $\lambda(\mathbf{x})$.
\end{enumerate}
 A Cox process is particularly suitable for inhomogeneous Poisson processes with intensity functions that vary over space, which is usually the case for environmental applications. Chapter 25 of \cite{handbook} states that Cox processes provide natural models when the point process in question arises as a consequence of environmental variation in intensity that cannot be described completely by available explanatory variables.

A particular kind of Cox process, the {\it log-Gaussian Cox process (LGCP)}, is defined with $\log\{\Lambda(\mathbf{x})\}$, a Gaussian spatial random field (\citenp{moller_et_al98}). 
LGCP models are effective and convenient in the sense that we can exploit the rich literature on spatial and spatio-temporal models for Gaussian random fields in geostatistics (\citenp{diggle_et_al13}).
In particular, stationary and nonstationary parametric mean and covariance functions, which have been developed in the geostatistical literature for both univariate and multivariate settings for spatial and spatio-temporal processes (e.g., \citenp{cressie_huang99,gneiting,stein05,apana_genton10,gneiting_et_al10,jun14}) can be used to model the stochastic intensity function. 
%Multivariate Cox processes, specifically, multivariate LGCPs, are convenient models for multivariate point patterns, as the problem then translates to the multivariate modeling of the multivariate intensity function.% (i.e. multivariate Gaussian random fields).

On the other hand, in the literature on multivariate point patterns with a multivariate stochastic intensity function, cross-covariance structures of the stochastic intensity functions have been quite limited. Suppose $\Lambda=(\Lambda_1,\ldots, \Lambda_m)$ denotes the multivariate stochastic intensity function. \cite[p. 126]{diggle} lets $\Lambda_1(\mathbf{x}) = \xi \cdot \Lambda_2 (\mathbf{x}) $ ($\xi>0$) for a bivariate case, which is too restrictive. \cite{moller_waagepetersen04} use a Linear Model of Coregionalization (\citenp{gelfand_et_al04}), which essentially writes each process as a linear combination of several common independent processes. 

Traditionally, most of spatial point process models were not developed for spherical domain. Common application examples for point processes in the literature concern spatial domains with sizes as small as an agricultural field or a small forest area, and the application domains are at most the size of a country (e.g., \citenp{schoenberg03,diggle_et_al05,shirota_gelfand17}). Recently, there have been developments in point pattern modeling on a global scale. \cite{robeson_et_al14} discussed how Ripley's $K$-function needs to be adjusted on spheres. \cite{lawrence_et_al16} discussed estimation and modeling of the K-function on spheres and applied {\it Thomas process models} on spheres to galaxies data set. \cite{moller_et_al18} developed {\it Determinantal point process models} on spheres. However, as far as the authors are aware, there has been little work on the LGCP models on a global scale, despite recent rapid developments on methods and models for geostatistical (continuous) spatial data on spheres (e.g., \citenp{heaton_et_al14,jun14,jeong_jun13,guinness_fuentes,porcu_et_al16}, see \cite{jeong_et_al17} for a review with more references). \cite{diggle_et_al13} list a series of application examples for LGCP models and all of these are assumed to be defined on $\mathbb{R}^2$. An R package, {\it lgcp} (\citenp{lgcp}), which provides a nice tool box for LGCP, does not consider cases for spatial patterns on a global scale. 

Our scientific interests, in this paper, are in understanding of the spatial patterns of rainfall occurrences for three rain types and how they are related to various atmospheric state variables. We present analysis of multivariate point patterns on a global scale through multivariate LGCP models. The nonstationary nature of occurrences of multiple rain types is dealt with by incorporating atmospheric state variables in the mean structure of the log of the stochastic intensity functions. The cross-covariance structure of multivariate (log) intensity functions for the three rain types is modeled by multivariate Mat\'{e}rn covariance function. We employ a Monte Carlo approximation of likelihood function for parameter estimation with the help of covariance approximation to deal with massive data. 

Although the rainfall data we use in this work is given on a gridded domain, we treat locations of grid points with rainfall as point patterns, rather than a lattice. This is reasonable given the high spatial and temporal resolution (we consider 0.5 degree, 6 hourly data) and the fact that global climate data is commonly a gridded product. In fact the data sets we use in this paper are originally from satellite and they are post processed on a high-resolution gridded domain. Rain pattern data have been analyzed using point process models in the literature. For instance, \cite{cowpertwait_et_al07} and \cite{kaczmarska_et_al14} applied point process models to deal with fine scale structure of rainfall process, although they focused on the temporal aspect of rainfall process. \cite{cowpertwait10} applied a spatio-temporal point process model for rainfall processes over the Rome region, Italy.

The rest of the paper is organized as follows. Section~\ref{sec:data} describes the data used for the analysis. Details in statistical models, inference, along with computational techniques for handling large data are given in Section~\ref{sec:method}. Section~\ref{sec:analysis} provides analysis results for modeling three rainfall types. The paper is concluded with some remarks in Section~\ref{sec:discussion}.

\section{Data}
\label{sec:data}

\subsection{Rainfall data}
\label{sec:rainfall}

The TRMM satellite (\citenp{kummerow_et_al98}) operated from late 1997 to early 2015 and yielded almost 17 years of continuous high-resolution measurements of the 3-dimensional structure of tropical and subtropical rainfall using its precipitation radar (PR). The PR had a footprint of 5 km at nadir and a 240 km swath width (these values were 4.3 km and 215 km before the 2001 altitude boost). About 2 million rain measurements were produced {\it per day}. The GPM satellite (\citenp{hou_et_al14}) has been operating since 2014 and has begun providing this information with the dual frequency precipitation radar (DPR) into the extratropics. 

 \begin{figure}[H] \centering
   { %\resizebox{15cm}{12.5cm}
     { {\includegraphics[angle=0,totalheight=6cm]{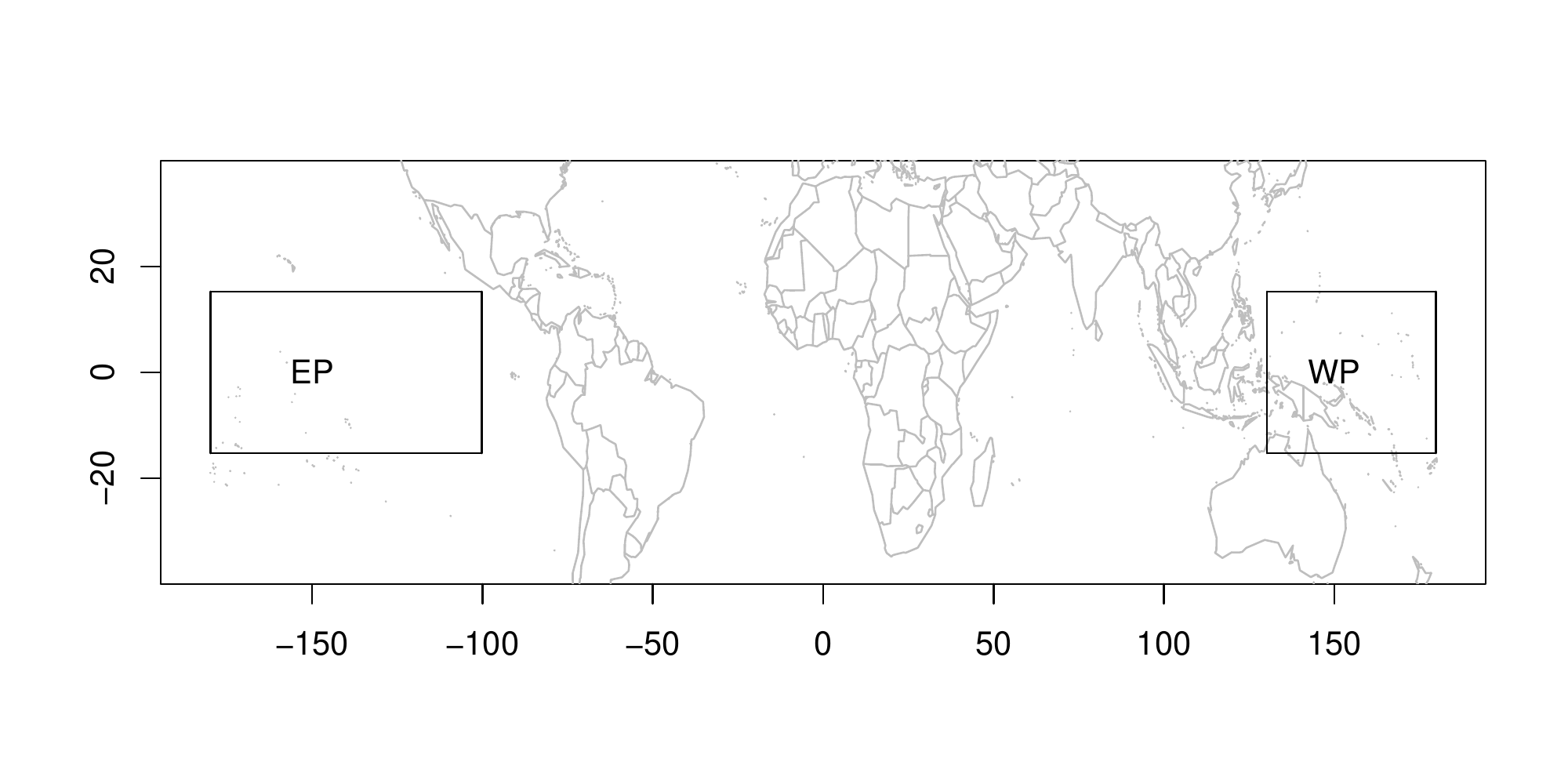}}}}
\caption{TRMM domain and the EP and WP regions}
\label{domain}
 \end{figure}

\begin{figure}[H] \centering
   { %\resizebox{15cm}{12.5cm}
     { {\includegraphics[angle=0,totalheight=15cm]{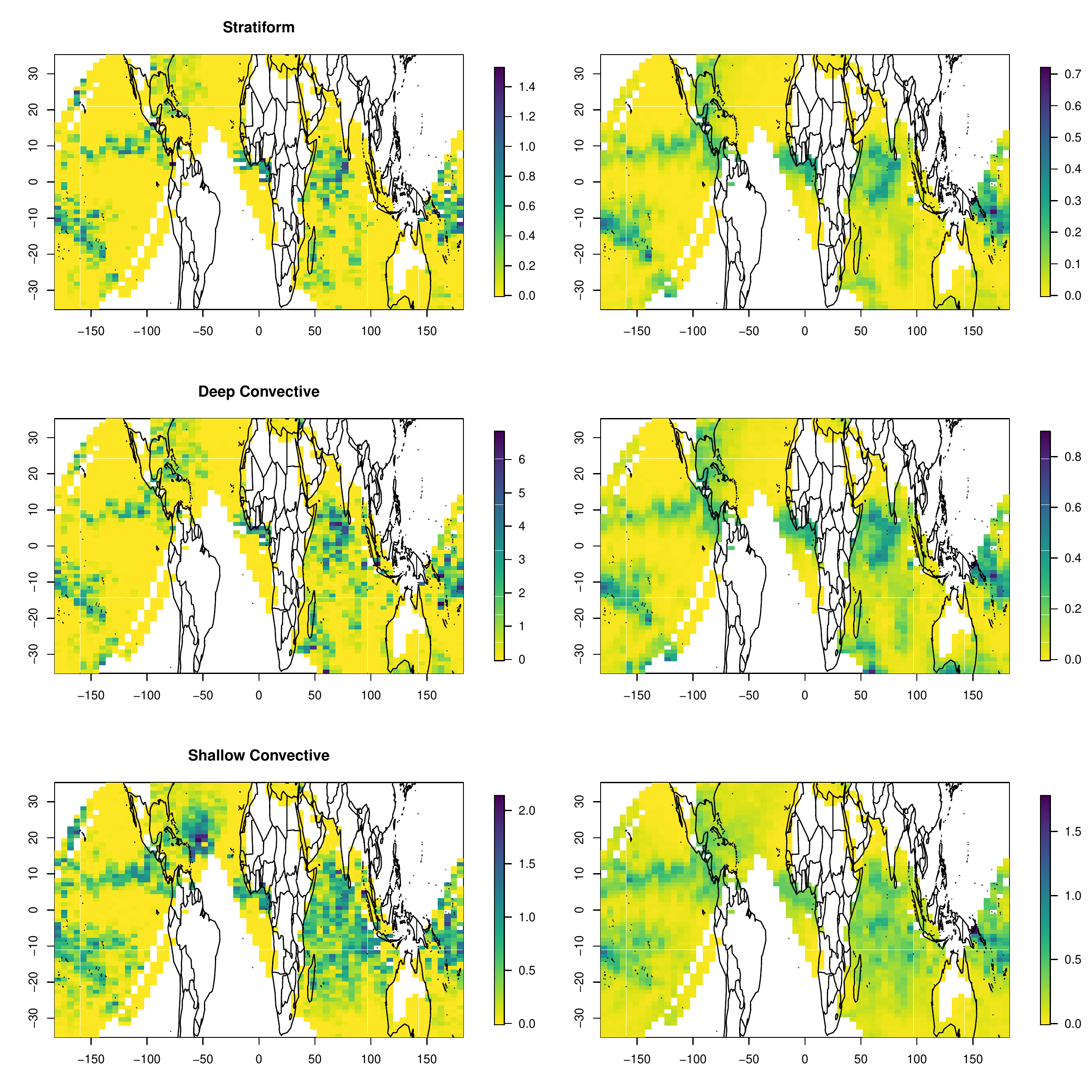}}}}
\caption{Rain intensity (mm/hr) (left) and estimated mean of log intensity process (right) for the three rain types (row-wise) during June 1-7, 2003}
\label{global}
 \end{figure}

We use Version 7 TRMM PR rainfall data for the first two weeks of June 2003 placed in 6-hourly, $0.5^{\circ}$ grids. 
June is the start of the Northern Hemisphere summer and is when rain begins to maximize in the tropical Pacific. The year 2003 was chosen because it was a neutral year for the El Nino-Southern Oscillation. Large rain changes happen across the Pacific when there is a warm El Nino event or a cold La Nina event. 
During this time period, only radar observations over the tropics and subtropics (from $35^{\circ}$ S to $35^{\circ}$ N) are available. Although this range does not cover the entire globe, it gives $360^{\circ}$ coverage in terms of longitude and thus it is inevitable to consider statistical models on a global scale (as opposed to models for planar domain). See Figure~\ref{domain} for the entire domain of TRMM data and Figure~\ref{global} for the spatial coverage of the TRMM path for the first week of June 2003. Note that there are approximately 16 orbits per day and that a $0.5^{\circ}$ grid will be visited by the PR 1-2 times per day at most (and often not at all) during a 6-hour period. The PR makes observations over both land and ocean, but we focus only on the ocean portions of the domain because rain type occurrence over land is strongly related to topography and the diurnal cycle of the sun (\citenp{ahmed_schumacher17}), thus complicating the statistical models.

The three rain types of interest are: {\it deep convective} (DC), {\it shallow convective} (SC), and {\it stratiform} (Str). Deep convection is associated with strong, intermittent rain and constitutes a large portion of rainfall over tropical land and oceans and the extratropical storm tracks. Stratiform cloud systems are associated with weaker, widespread rainfall that can either form as a result of deep convective clouds, as is common in tropics, or from large-scale lifting as found in fronts at higher latitudes (\citenp{houze04}). Convective rain in general can be separated into shallow and deep, where all of the shallow convective rain forms from warm rain processes and cloud tops do not exceed the $0^{\circ}$ C height level (\citenp{schumacher_houze03}). Deep convective cloud tops often exceed 10 km and cold rain processes play an important role in overall intensity and rain production. Shallow convection often occurs outside of the heavy rain regions in the tropics, unlike deep convection. All three rain types are differentiated in the PR observations using texture and height information (\citenp{awaka_et_al97,funk_et_al13}). 

\subsection{Atmospheric state variables}
\label{sec:state}

The atmospheric state variables that will help describe the rainfall distributions are generated by a global climate model that assimilates data to provide a dynamically consistent set of fields constrained by observations. We use NASA's Modern-Era Retrospective analysis for Research and Applications, Version 2 (MERRA-2; \citenp{molod_et_al15}). We use 6-hourly data at $2/3^{\circ} \times 1/2^{\circ}$  horizontal grid resolution. The reanalysis fields utilized are temperature, humidity, horizontal winds, and surface latent heat flux. These variables are interpolated to a common horizontal grid with $0.5^{\circ}$ spatial resolution to match the spatial resolution of the gridded TRMM rainfall data every 6 hours. Care is taken to preserve the predictive temporal relationships (for instance, if atmospheric state variables are observed at time 00 UTC, then rainfall data is accumulated from 00 UTC to 06 UTC). This permits the attribution of causal interpretations to any statistical relationships that are identified.
 
Some atmospheric state variables, such as temperature, humidity, and horizontal winds, are given for multiple vertical levels from the reanalysis. One of the main scientific interests in this paper is to relate the vertical profile of these state variables with the rainfall data near the surface. Atmospheric scientists often use a technique called {\it Empirical Orthogonal Function (EOF)} decomposition. This is essentially the same as Principal Component (PC) analysis and each EOF is a vector of weights (or loadings) for each level of a PC. That is, let $t(\mathbf{s},h)$ denote the temperature value at spatial location $\mathbf{s}$ and vertical height (i.e. pressure) $h$, and temperature is observed at $r$ pressure levels, $h_1,\ldots,h_r$. If $i$th PC ($i=1,\ldots,r$) is expressed as $t_i(\mathbf{s})= a_1 t(\mathbf{s},h_1) + \cdots a_r t(\mathbf{s},h_r)$, the corresponding EOF, $E_i(\mathbf{s})$, is given by $(a_1,\ldots,a_r)$.   

 \begin{figure}[H] \centering
   { %\resizebox{15cm}{12.5cm}
     { {\includegraphics[angle=0,totalheight=12cm]{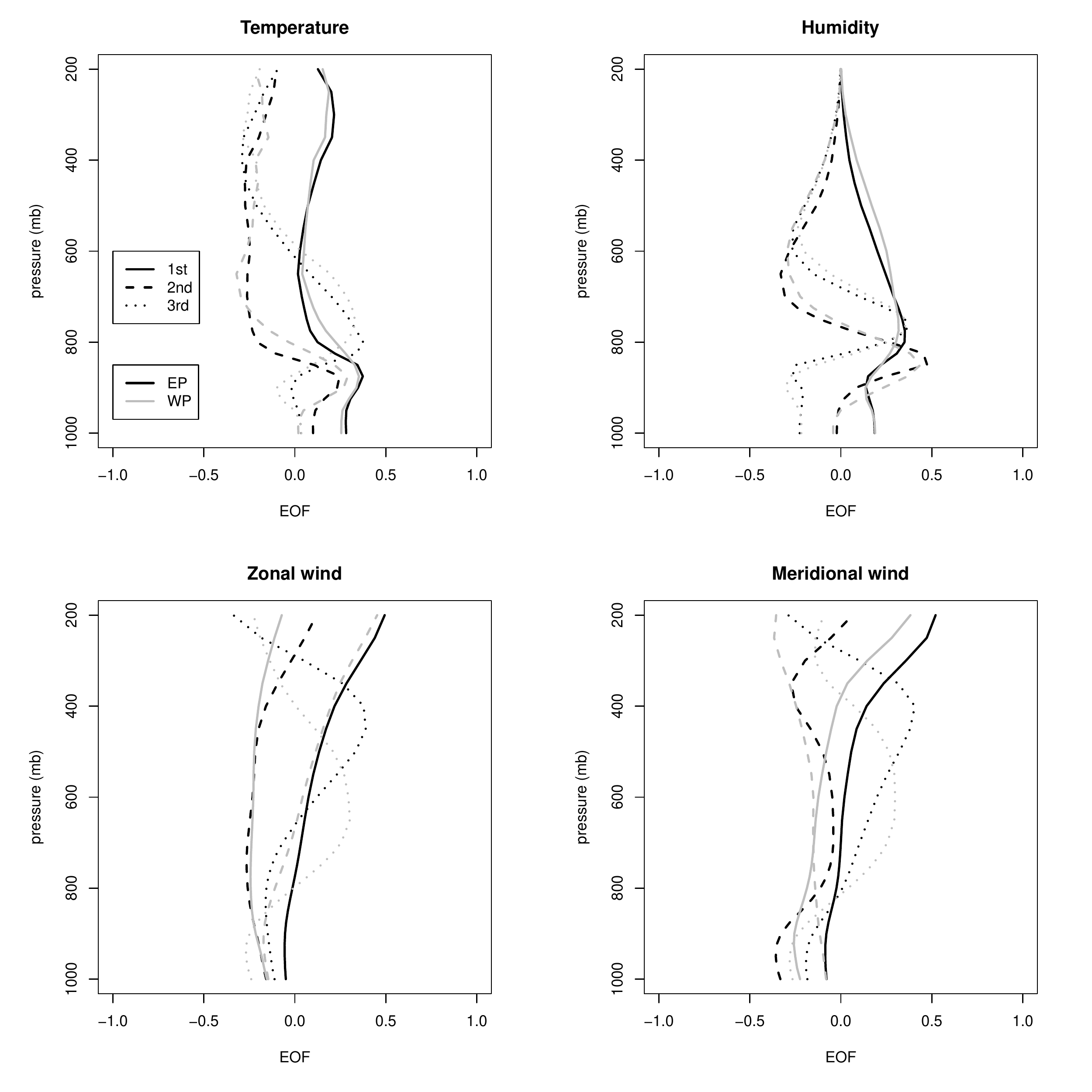}}}}
\caption{First three EOFs of temperature, humidity, and wind data over the Eastern and Western Pacific regions}\label{EOF}
 \end{figure}

Figure~\ref{EOF} shows first three EOFs of temperature (corresponding to $t_1,t_2,t_3$), humidity ($q_1,q_2,q_3$), zonal (east-west) winds ($u_1,u_2,u_3$) and meridional (north-south) winds ($v_1,v_2,v_3$) over the Eastern Pacific (EP) and Western Pacific (WP) domains from Figure~\ref{domain}. Particular features of note are the differing heights and depth of the inversion (i.e. when temperature increases with height) at low levels in each temperature profile, the strong drying or moistening at mid levels in the humidity profiles, and the strength and direction of the winds near the surface and in the upper atmosphere. Negative zonal values indicate easterly winds (i.e., winds from the east) and positive zonal values indicate westerly winds (i.e., winds from the west). Similarly, negative meridional values indicate northerly winds (or winds from the north) and positive meridional values indicate southerly winds (or winds from the south). Temperature inversions are generally detrimental to convection unless they can be broken through (e.g., by daytime surface heating) allowing convection to attain great strength. A dry mid atmosphere is considered detrimental to both deep convective and stratiform rain, but is commonly associated with shallow convection (e.g., \citenp{jensen_et_al06}). The relationship between wind and rainfall is often linked to the change of wind speed and/or direction with height (i.e., wind shear), which is further discussed below.

Additional horizontally-varying state variables considered in the study are three definitions of vertical wind shear ($ls$, $dp$, and $dds$), surface latent heat flux ($lh$), and latitude ($lat$). If $u[z]$ and $v[z]$ denote zonal and meridional wind speeds at pressure level $z$, then shear variables are defined in the following way:
\begin{align*}
&ls=\sqrt{(u[900]-u[700])^2 + (v[900]-v[700])^2}& \mbox{(low-level shear)},\\
&dp=\sqrt{(u[900]-u[300])^2 + (v[900]-v[300])^2}& \mbox{(deep shear)},\\
&dds=u[300]-u[900]& \mbox{(deep directional shear)},
\end{align*}
where $900$ denotes the $900$ mb pressure level and so on. The three vertical shear variables are meant to highlight different mechanisms in the atmosphere that promote convective and stratiform rain production. For example, low-level shear typically helps initiate convective cells (\citenp{rotunno_et_al88}), while deep shear is thought to assist the formation of stratiform rain regions (\citenp{li_schumacher11}). Deep directional shear represents situations when low-level zonal winds are going the opposite direction of the upper level zonal winds, which would cause the low-level convective cloud bases to rapidly move in a different direction than any upper level cloud, potentially impacting the occurrence of deep convective and stratiform rain.

\section{Statistical method}
\label{sec:method}

The statistical challenges for this work come from the following: (i) multivariate spatial point patterns, (ii) point pattern models on a global scale, (iii) statistical inference for such point pattern models, and (iv) computational difficulties due to large number of data points.

\subsection{LGCP model for multivariate point patterns on a sphere}

Suppose $X$ is a spatial point pattern on the surface of a sphere, $\mathcal{S}^2 \subseteq \mathbb{R}^3$ and $Y$ is a Gaussian random field on $\mathcal{S}^2$. Let the mean and covariance function of $Y$ be 
$$m(\mathbf{s})= E Y(\mathbf{s}),~~C(\mathbf{s}_1, \mathbf{s}_2)= \mbox{Cov} \{Y(\mathbf{s}_1), Y(\mathbf{s}_2)\},$$ for $\mathbf{s}, \mathbf{s}_1, \mathbf{s}_2 \in \mathcal{S}^2$. We assume $Y$ is a log-Gaussian Cox process driven by $\Lambda=\exp(Y)$.  
%It is easy to see that $\rho(\mathbf{s})= \exp\{m(\mathbf{s})+C(\mathbf{s}, \mathbf{s})/2\}$ and $g(\mathbf{s}_1, \mathbf{s}_2)= \exp\{C(\mathbf{s}_1, \mathbf{s}_2)\}$. (definition)

Depending on the structure we give to $m$ and $C$, the resulting LGCP may have various properties. For example, one might give a simple structure by assuming $m(\mathbf{s})=\mu$ and $C(\mathbf{s}_1, \mathbf{s}_2)= C_0(d_{12})$ with $d_{12}$ a great circle distance between $\mathbf{s}_1$ and $\mathbf{s}_2$ on $\mathcal{S}^2$. 
%In that case, resulting Poisson process has $\rho(\mathbf{s})= \exp\{\mu+C_0(0)/2\}$ and $g(\mathbf{s}_1, \mathbf{s}_2)= \exp\{C_0(d_{12})\}$. 
Note that \cite{chakraborty_et_al11} used this structure except they did not assume a sphere as their spatial domain. On the other hand, one might assume that the mean structure of $Y$ varies over the surface of a sphere with an isotropic covariance structure for $Y$, or assume a constant mean structure but a nonstationary covariance structure for $Y$. %In that case, $g$ is nonstationary, and $\rho$ may or may not vary over space depending on how variance of $Y$ varies over space.

For the application in this paper, we utilize atmospheric state variables (described in Section~\ref{sec:state}) to account for the nonstationary mean structure of the log of the intensity process, $Y_i$ ($i=1,2,3$ for each rain type). Further, the natural clustering of points for the locations of rainfall will be dealt with through an isotropic Mat\'{e}rn covariance model. %: $\mbox{Cov}\{Y_i(\mathbf{s}_1), Y_i(\mathbf{s}_2)\}= \frac{\alpha}{2^{\nu-1}\Gamma(\nu)} (\frac{d_{12}}{\beta})^{\nu} \mathcal{K}_{\nu}(\frac{d_{12}}{\beta})$. 
Indeed, we use the parsimonious version of the Mat\'{e}rn covariance model originally introduced by \cite{gneiting_et_al10}. We not only are able to estimate the contribution of each state variable to the occurrence of rainfall for each type, but are able to estimate the cross-correlation of pairs of rain types through this multivariate LGCP model.  

Although the multivariate Mat\'{e}rn models originally introduced by \cite{gneiting_et_al10} are not developed for processes on spheres, recently \cite{gneiting11} (for univariate) and \cite{porcu_et_al16} (for multivariate) showed conditions on parameters to ensure positive definiteness of multivariate Mat\'{e}rn model on spheres. There are total of 6 smoothness parameters for the Mat\'{e}rn model used for this application: $\nu_i$ and $\nu_{ij}$ for $i,j=1,\ldots,3$. We fix all of them equal to 0.5, and the resulting model satisfies the condition for positive definiteness according to \cite{porcu_et_al16}. 
We could extend the model further by employing a nonstationary covariance function for modeling $Y_i$'s, such as models introduced in \cite{stein05_cises} and \cite{jun14}. 

\subsection{Monte Carlo likelihood approximation}

Commonly, statistical inference for point process models is done through either a moment-based method or likelihood. The commonly used moment-based method, called {\it minimum contrast}, finds parameter estimates by minimizing the squared difference between the empirical and theoretical versions of Ripley's $K$-function. See chapter 19 of \cite{handbook} for more details.  Recently, \cite{robeson_et_al14} pointed out that the $K$ function needs to be adjusted for point patterns on spheres. Nevertheless, similar to the least squares method for variogram estimation, moment-based methods for inference for point pattern models are known to be less efficient compared to likelihood-based methods (\citenp{diggle}).

The likelihood for a LGCP model with data  $X=\{x_i \in A: i=1,\ldots,n\}$ defined on a spatial domain $W$ is given as (\citenp{diggle_et_al13}) 
\begin{align}L(\theta;X) = P(X|\theta)= \int_{\Lambda} P(X,\Lambda|\theta) d \Lambda = E_{\Lambda|\theta} ( L^*(\Lambda;X)) \label{likelihood}
\end{align} 
with \begin{align}
l^*(\Lambda;X)= \log{L^*(\Lambda;X)}= \sum_{i=1}^n \Lambda(x_i) - \int_W \Lambda(u)du. \label{likelihood1}
\end{align}
The main obstacle in performing maximum likelihood estimation for LGCP models has been that the evaluation of \eqref{likelihood} involves integration over the infinite-dimensional distribution of $\Lambda$.
For a given realization of $\Lambda$, the integral term in \eqref{likelihood1} on the surface of a sphere adds further computational complication.

%\subsection{poisson counts}

%For some applications, data sets are given as aggregated counts on (often regular) cells over space. (explain an example)
% That is, instead of random locations for each event, sum of the events for each grid cell $s_{ij}$, $N_{ij}$, is given. In this case, it is natural to consider $N_{ij} \sim Poisso\mathcal{N}(\Lambda_{ij})$ with $\Lambda_{ij}= \int_{s_{ij}} \lambda(\mathbf{x})d\mathbf{x}$ and $\Lambda_{ij}$ is approximated as $|s_{ij}| \lambda(s_{ij})$ \citenp{waagepetersen04}. Here $|s_{ij}|$ denotes the area of the cell $s_{ij}$ on the sphere.

%(discuss example)

%\subsection{Monte Carlo likelihood}

 A natural alternative to integrating the likelihood function over the stochastic intensity function (i.e., calculating the expected value with respect to the stochastic likelihood function in \eqref{likelihood}) is through Monte Carlo approximation. That is, the expectation is approximated by an empirical average over a simulated realization. For simulated realizations of $\Lambda$, $\lambda^{(j)}= \{\lambda^{(j)}(\mathbf{s}_k):k=1,\ldots,N\}$, $j=1,\ldots, s$, with a finite ``grid'' points $\mathbf{s}_1,\ldots,\mathbf{s}_N$ that cover the spatial region of interest, one may approximate \eqref{likelihood} with
\begin{align}
L_{MC}(\theta)= \frac{1}{s} \sum_{j=1}^s L(\theta; X, \lambda^{(j)}). \label{MC}
\end{align}
Here, we need to consider finite grid points since we cannot simulate random fields continuously over space. The accuracy of the approximation in \eqref{MC} depends on $s$. 
The idea of a Monte Carlo approximation of likelihood for Cox processes has not been utilized much in the past mainly because of its computational intensiveness. For a LGCP model, such likelihood approximation requires a large number of simulated Gaussian random fields over dense grid points. 

With the recent development of technology and computing power, however, simulating a large number of Gaussian random fields over dense grid points becomes more doable. An R package, {\it RandomFields} (\citenp{randomfields}), provides tools to simulate Gaussian random fields over a large number of locations over Euclidean domain as well as spheres. Computational techniques for approximating likelihood (\citenp{stein_chi_welty04,fuentes07}) or composite likelihood (\citenp{cox_reid04}) may not be directly applicable since we need to simulate random fields, rather than calculate likelihood values.

\subsection{Covariance approximation}
\label{sec:fsa}

Suppose one needs to simulate a Gaussian random field over $\mathbf{s}_1,\ldots,\mathbf{s}_N$ for $s$ many times. Let $\mathbf{Y}= \{Y(\mathbf{s}_1), \ldots, Y(\mathbf{s}_N)\} \sim \mathcal{N}(\boldsymbol{\mu}, \boldsymbol{\Sigma})$ with $\boldsymbol{\Sigma}$ an $N \times N$ covariance matrix. 
For computationally efficient simulation of a large number of Gaussian random fields over a large number of locations (that is, large $N$), we will use the idea of a predictive process (PP) model (\citenp{banerjee2008gpp}) in a non-Bayesian context. 
PP models have been proven to provide computationally efficient tools for dealing with Gaussian random fields observed over a large number of spatial locations. However, their weakness in dealing with small-scale spatial variations has also been observed (e.g., \citenp{sang_et_al11,stein14}). 
We will use a modified version of PP proposed in \cite{sang_et_al11,sang+huang} to account for the large-scale, as well as small-scale, variation of each Gaussian random field. 

We approximate $\boldsymbol{\Sigma}$ by  
\begin{align}
\boldsymbol{W} = \mathbf{A} \mathbf{R}^{-1} \mathbf{A}^T + \mathbf{V}, \label{cov_approxi}
\end{align}  
a part resulting from a PP model and an approximated remainder. Here, we introduce a set of knots, $\mathbf{u}_1, \ldots, \mathbf{u}_m$, for $m << N$ that cover the entire domain. Then, $\mathbf{R}= \mbox{Var}\{\mathbf{Y}(\mathbf{u})\}$, $\mathbf{u}=\{\mathbf{u}_1, \ldots, \mathbf{u}_m\}$ and $\mathbf{A}=\mbox{Cov}\{\mathbf{Y}(\mathbf{s}), \mathbf{Y}(\mathbf{u})\}$, $\mathbf{s}=\{\mathbf{s}_1,\ldots, \mathbf{s}_N\}$.  
Furthermore, the ``remainder'' covariance matrix, $\boldsymbol{\Sigma} - \mathbf{A} \mathbf{R}^{-1} \mathbf{A}^T$, after subtracting the covariance matrix for a PP model, is approximated by the ``block independent'' adjustment (c.f. adjustment using taper functions as in \cite{sang+huang}). That is, the remainder matrix is approximated by a block diagonal matrix, denoted by $\mathbf{V}$.

We then simulate multiple Gaussian random fields based on the approximation in \eqref{cov_approxi}. Let $\tilde{\mathbf{S}}$ be a $s \times N$ matrix with $s$ being many simulated random fields over $N$ locations, $\mathbf{S}_0$ a $s \times m$ matrix with elements from {\it iid} $\mathcal{N}(0,1)$, and $\mathbf{S}_1$ a $s \times N$ matrix with elements from {\it iid} $\mathcal{N}(0,1)$ (each element in $\mathbf{S}_1$ is independent of elements in $\mathbf{S}_0$). 
Then, we write 
\begin{align}
\tilde{\mathbf{S}} = \mathbf{S}_0 \mathbf{B} + \mathbf{S}_1 \mathbf{U}_V, \label{simulation}
\end{align}
 where $\mathbf{B} = (\mathbf{U}_R^{-1})^T \mathbf{A}^T$, and $\mathbf{U}_R$ and $\mathbf{U}_V$ are upper triangular matrices resulting from Cholesky decomposition of $\mathbf{R}$ and $\mathbf{V}$, respectively (that is, $\mathbf{R}= \mathbf{U}_R^T \mathbf{U}_R$ and $\mathbf{V}=\mathbf{U}_V^T \mathbf{U}_V$). To find $\mathbf{B}$, instead of inverting $\mathbf{U}_R$, we use $$\mathbf{B}=(\mathbf{U}_R^{-1})^T \mathbf{A}^T = (\mathbf{U}_R^T)^{-1} \mathbf{A}^T \Leftrightarrow \mathbf{U}_R^T \mathbf{B} = \mathbf{A}^T$$ and solve for $\mathbf{B}$ efficiently using forward solve algorithm. 
The Cholesky decomposition of $\mathbf{V}$ can also be done efficiently accounting for the fact that $\mathbf{V}$ is a block diagonal matrix. That is, one needs to perform multiple Cholesky decomposition of each block of $\mathbf{V}$ to reduce computation significantly.
See Appendix for a short proof to show that the covariance matrix of each column of $\tilde{\mathbf{S}}^T$ equals the approximation of $\boldsymbol{\Sigma}$ given in \eqref{cov_approxi}.

\section{Applications}
\label{sec:analysis}

We use TRMM satellite radar data as well as MERRA-2 atmospheric state variable data for June 2003 as described in Section~\ref{sec:data}. Understanding rain distributions during the summer months in the tropical Pacific is especially important because of the strength of the Intertropical Convergence Zone (ITCZ), a region of enhanced convection at the intersection of the trade winds, during these months and the overall importance of the tropical Pacific to the onset and evolution of El Ni\~{n}o events. We first consider rainfall data over the EP and the WP regions during the first two weeks of June (Section~\ref{sec:local}). Then we analyze the global data for the first week of June (Section~\ref{sec:global}). 
Each atmospheric variable is standardized so that its spatial mean equals zero and its standard deviation equals one.

Let $Y_i$ be the log transformed intensity process for $i$th rain type ($i=1,2,3$). For the mean structure of the log transformed intensity process, $Y_i$, we write
\begin{align}
E\{Y_i(\mathbf{s})\} = \eta_{0,i} + \eta_{1,i}~ t_1(\mathbf{s})+\eta_{2,i} ~t_2 (\mathbf{s}) + \cdots + \eta_{17,i}~ lat(\mathbf{s}).\label{linear-model}
\end{align}
For the covariance structure of $\mathbf{Y}=(Y_1,Y_2,Y_3)$, we use a trivariate version of Mat\'{e}rn covariance function (\citenp{gneiting_et_al10,porcu_et_al16}). For model parsimony, we give a common spatial range parameter ($\beta$, in spherical distance) and focus on estimating the cross-correlation between the three rain types ($\rho_{ij}$, $i,j=1,2,3$). Indeed, we tried to fit the model separately for each rain type and found that estimates for spatial range parameter ($\beta$) do not vary much across different rain types. Note that with common spatial range parameter, the full cross-covariance matrix reduces to a Kronecker product of $3\times 3$ cross-covariance matrix and a univariate spatial correlation matrix, which is again, guaranteed to be positive definite. 
  
\subsection{EP vs WP}
\label{sec:local}

The EP region covers a longitude range from $180^{\circ}$ W to $100.25^{\circ}$ W and the WP region covers a longitude range from $130.25^{\circ}$ E to $180^{\circ}$ E (Figure~\ref{domain}). Both regions cover a latitude range of $15.25^{\circ}$ S to $15.25^{\circ}$ N.
We set $s=10000$ for the Monte Carlo approximation of likelihood function (as in \eqref{MC}). We tried various values of $s$ and the results did not change significantly as long as $s$ is reasonably large (e.g., $s \geq 5000$). Analytic calculation of the integral in \eqref{likelihood1} is not possible and thus the integral term is approximated by a Riemann sum (using 5000 many terms). Note that we do not use covariance approximation for this local analysis as we can afford to use full covariance matrix for the simulation of Gaussian random fields. We let $N$ equal the total number of grid pixels for the TRMM data in the EP or WP region.

Table~\ref{covariate} shows coefficient estimates for the atmospheric state variables for the log intensity processes. %Covariance parameter estimates for the trivariate Mat\'{e}rn model, including cross-correlation estimates for pairs of rain types, are given in Table~\ref{matern-parameter}. 
The best predictor for rain type occurrence is humidity, consistent with our physical expectation and previous statistical studies (e.g., \citenp{chen_et_al17}). In particular, the first humidity EOF ($q_1$) indicates a moister atmosphere throughout the depth of the troposphere, which is strongly conducive to rain production, especially DC and Str rain types. The second humidity EOF ($q_2$) indicates a drier mid-troposphere (e.g., 600-700 mb), which hinders deep convective cloud growth making it a better predictor for SC rain.

\begin{table}
\begin{center}
\caption{Estimates for coefficients of the linear model in \eqref{linear-model}. See Section~\ref{sec:state} for definitions of predictors}
\label{covariate}
\begin{tabular}{|c||r|r|r||r|r|r||r|r|r|}
\hline
&\multicolumn{3}{|c||}{EP}&\multicolumn{3}{|c||}{WP}&\multicolumn{3}{|c|}{Global}\\
\cline{2-10}
Predictor&Str&DC&SC&Str&DC&SC&Str&DC&SC\\
\hline
%$\eta_0$&$-$3.640&$-$3.159&$-$1.552&$-$2.573&$-$2.285&$-$1.518&$-$3.294&$-$2.417&$-$1.264\\
$t_1$& 0.61&  0.45&  0.12&0.31&0.25&0.03&$-$1.53&$-$1.64&$-$1.00\\
$t_2$&  0.80&  0.63&  0.31&0.23&0.27&0.04&0.32&0.44&0.12\\
$t_3$& $-$0.43& $-$0.29 &$-$0.29&$-$0.11&$-$0.13&$-$0.10&0.01&0.23&$-$0.14\\
\hline
$q_1$& 1.15&  1.18 & 0.87&0.88&0.76&0.46&2.67&2.24&1.63\\
$q_2$& 0.24&  0.22&  0.39&0.09&0.11&0.22&$-$0.23&$-$0.20&$-$0.35\\
$q_3$& $-$0.08& $-$0.06& $-$0.08&$-$0.06&$-$0.05&$-$0.05&$-$0.17&$-$0.13&0.17\\
\hline
$u_1$& 0.10&  0.15&  0.11&0.00&0.00&$-$0.02&$-$0.58&$-$0.17&0.04\\
$u_2$& $-$0.14& $-$0.15& $-$0.07&$-$0.01&$-$0.07&$-$0.06&$-$0.16&0.06&$-$0.14\\
$u_3$&  0.05&  0.03& $-$0.07&$-$0.03&$-$0.03&0.03&0.04&0.02&0.04\\
\hline
$v_1$& $-$0.03& $-$0.04&  0.04&$-$0.01&0.01&0.06&$-$0.02&$-$0.03&$-$0.09\\
$v_2$& $-$0.14& $-$0.11& $-$0.08&0.11&0.08&0.06&0.06&0.04&0.09\\
$v_3$& 0.09&  0.07&  0.09&$-$0.14&$-$0.10&$-$0.06&0.04&0.01&$-$0.02\\
\hline
ls& $-$0.02& $-$0.05& $-$0.10&$-$0.14&$-$0.14&$-$0.08&0.08&0.02&$-$0.06\\
dp& $-$0.07&  0.00&  0.01&$-$0.01&$-$0.01&$-$0.08&$-$0.15&$-$0.14&0.05\\
dds& $-$0.02& $-$0.09& $-$0.10&$-$0.02&$-$0.05&0.11&$-$0.16&0.17&0.03\\
lh&  0.03&  $-$0.01 & $-$0.01&$-$0.05&$-$0.05&$-$0.01&0.05&0.16&0.21\\
lat &$-$0.04& $-$0.09& $-$0.12&0.09&0.08&0.11&$-$0.12&$-$0.07&0.21\\
\hline
\end{tabular}
\end{center}
\end{table}

The next best predictor for rain type occurrence is temperature. The second temperature EOF ($t_2$) is warm at low levels and then rapidly cools around 800 mb. This creates an unstable temperature profile that promotes a deep convecting atmosphere. The first temperature EOF ($t_1$) has a similar structure to $t_2$ except that it is warmer at low levels than upper levels and does not cool quite as rapidly at 800 mb. The third temperature EOF ($t_3$) indicates a strong inversion around 800 mb, which would damp convective cloud growth and explains the negative coefficients in Table~\ref{covariate}. These temperature EOFs are all good predictors for DC and Str but are weaker predictors for SC rain.

More generally, SC rain tends to have weaker or (sometimes even opposite) relationship compared to DC and Str rain for most of the predictors in Table~\ref{covariate}. This result is physically consistent with the fact that stratiform rain forms from deep convection in the tropics, while shallow convective rain can occur outside of regions of deep convection. Table~\ref{matern-parameter} further highlights these rain type relationships with cross-correlation values greater than or equal to 0.95 for Str and DC rain but negative or near zero for Str and SC rain.

\begin{table}
\begin{center}
\caption{Estimates for covariance parameters for trivariate Mat\'{e}rn covariance function. The unit for $\beta$ (spatial range in spherical distance) is km. Cross-correlation $\rho_{12}$ is between Str and DC, $\rho_{13}$ between Str and SC, and $\rho_{23}$ between DC and SC} %(beta for global seem to fall in local maxima, for smaller fit. larger fit going on)}
\label{matern-parameter}
\begin{tabular}{|l|r|r|r|}
\hline
&EP&WP&Global\\
\hline
$\beta$&{1465.57}&{1211.97}&{1096.63}\\
\hline
{$\rho_{12}$} & {0.95}&{0.99}&{0.99}\\
$\rho_{13}$&$-0.11$&$-$0.01&$-$0.01\\
$\rho_{23}$&{0.18}&{$-$0.01}&$-$0.01\\
\hline
\end{tabular}
\end{center}
\end{table}

While the predictors related to wind, surface latent heating, and latitude have lower coefficients than temperature and humidity in Table~\ref{covariate}, they still provide information about rain type occurrence and its regionality. For example, zonal wind variations are better predictors in the EP, while low-level shear is a better predictor in the WP. Outside of temperature and humidity, latitude is the strongest predictor for SC rain. Table~\ref{matern-parameter} also indicates similar spatial range ($\beta$, in spherical distance) between the two regions.%, which may be partly explained by the fact that the WP is warmer and has more homogeneous sea surface temperatures than the EP.

\begin{figure}[H] \centering
   { %\resizebox{15cm}{12.5cm}
     { {\includegraphics[angle=0,totalheight=14cm]{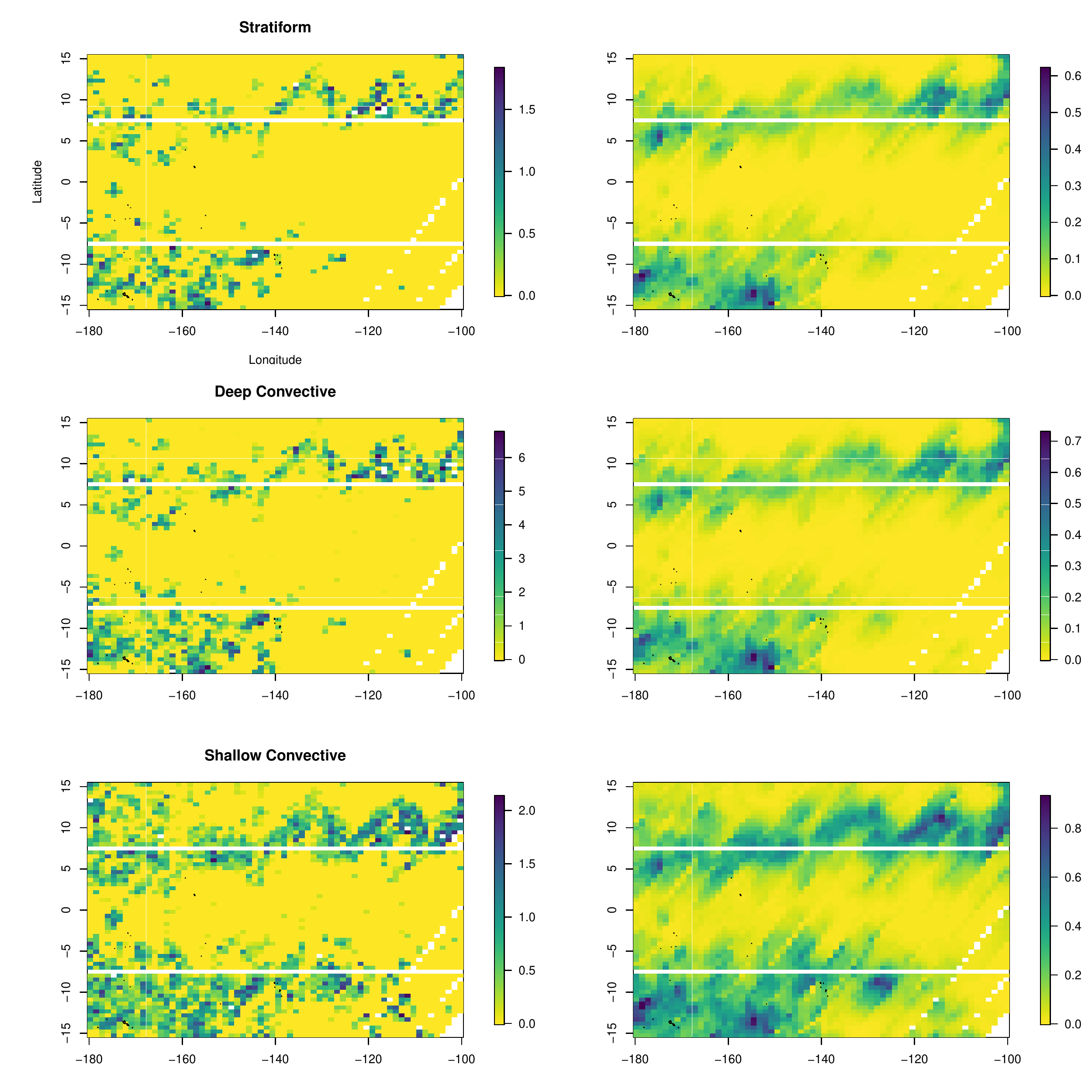}}}}
\caption{Rain intensity (mm/hr) (left) and estimated mean of log intensity process (right) for the three rain types (row-wise) over the EP region during June 1-14, 2003}
\label{figure-EP}
 \end{figure}

\begin{figure}[H] \centering
   { %\resizebox{15cm}{12.5cm}
     { {\includegraphics[angle=0,totalheight=14cm]{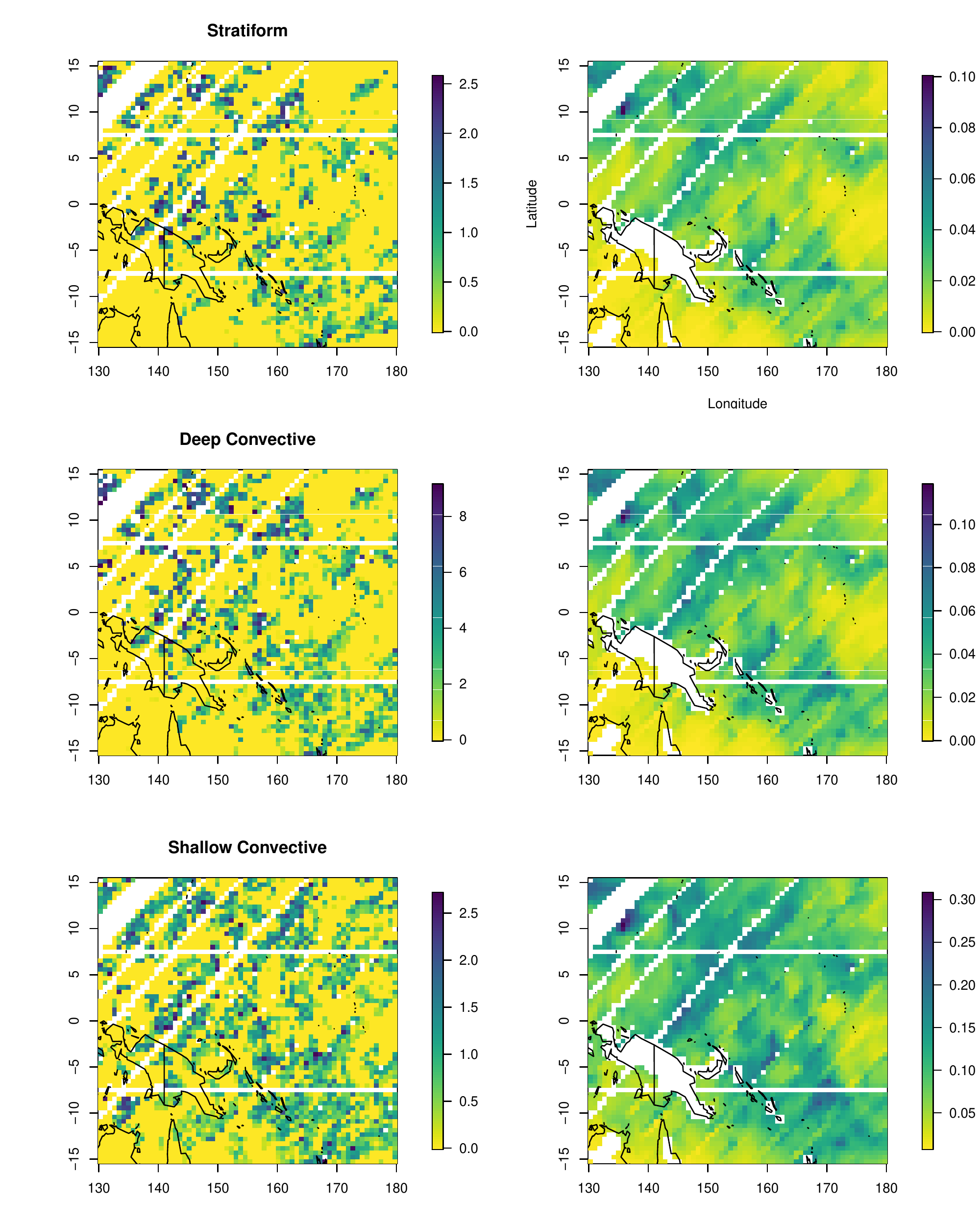}}}}
\caption{Same as Figure~\ref{figure-EP} over the WP region}
\label{figure-WP}
 \end{figure}

Figures \ref{figure-EP} and \ref{figure-WP} show map comparisons between the observed rain intensity
and the estimated mean structure of the log intensity process for each
rain type. In the EP, Str and DC are narrowly confined to the warm ocean
regions north and south of the equatorial cold tongue, while SC is more
spatially distributed (Figure~\ref{figure-EP}). In the WP, rain occurrence for each
type is more evenly distributed across the domain because of the
generally warm waters in the western tropical Pacific (Figure~\ref{figure-WP}). Even
though we only take occurrences of rain for each type and we did not
take into account the actual rain intensity (or rain rate) in our
analysis, overall spatial patterns of the observed rain intensity (left columns of the figures) and estimated mean log intensity (right columns of the figures) are strikingly similar. For both regions and all rain
types, the original rain rate data is quite noisy and the estimated mean
field for the log intensity process appears much smoother. This is
because we only display the estimated mean field, as given in \eqref{linear-model}.

\subsection{Global analysis}
\label{sec:global}

We now perform a multivariate analysis for spatial point patterns for the three rain types over the entire TRMM domain.% ($35^{\circ}$ N to S, still limited to only ocean regions). 
We consider TRMM PR data for June 1 to June 7 in 2003, as well as corresponding atmospheric state variables (note that EOFs are calculated again for the larger domain). Even with only one week of data, the number of grid pixels covered during the period is around 75,000 (and thus $N \approx 75,000$). Therefore, we need to approximate covariance matrices as described in Section~\ref{sec:fsa}. 

%As shown in Figure~\ref{global}, TRMM tracks for the time period considered cover mostly over the ocean.%ut there are some land area that are covered.
Similar to the local analysis in Section~\ref{sec:local}, we express the log intensity function of the LGCP as a linear combination of all the atmospheric state variables. We also use a parsimonious version of the trivariate Mat\'{e}rn covariance function similar to the EP/WP analysis.
We let $s=400$ for the Monte Carlo approximation of likelihood. For the full-scale approximation, we use $m=100$ and let the size of each block matrix in $\mathbf{V}$  be equal to $100 \times 100$.

The right columns of Tables~\ref{covariate} and \ref{matern-parameter} show estimated coefficients for the mean of log intensity process as well as covariance parameters for the log intensity process over the tropical and subtropical ocean regions. It is interesting to note that, while the estimated coefficients for EP and WP are similar, estimated coefficients for the global analysis are somewhat different.% Furthermore, estimated spatial range parameter ($\beta$) for the global analysis is larger compared to the local analysis. 
 However, humidity and temperature remain the best predictors for all rain types and SC coefficients tend to be less than or of opposite sign than the Str and DC coefficients.
Cross-correlation estimates for the global analysis are similar to those from the local analysis.
The likelihood function turns out to be quite flat for the spatial range parameter. 
Figure~\ref{global} reiterates the capability of the statistical analysis to accurately capture not only each rain type occurrence, but their rain rates as well.

\section{Concluding remarks}
\label{sec:discussion}

We demonstrated that LGCP models can be applied to local as well as global spatial point pattern data in a multivariate setting. We applied Monte Carlo approximations to log likelihood functions, and for the global analysis, we exploited covariance approximation methods to ease computational difficulties due to massive data. We were able to tease out scientifically interesting connections between rainfall occurrences and atmospheric state variables as well as cross-correlation between multiple rain type occurrence patterns. 

We have shown that profiles of humidity and temperature, and even single-level variables, can predict the occurrence and intensity of rainfall in the tropics separated into deep convective, stratiform and shallow convective components. These three rain types are the building blocks of tropical cloud systems at multiple time and space scales (\citenp{mapes_et_al06}), so the ability to predict rain type characteristics from environmental observations using statistical models supports the feasibility of parameterization of organized convective systems in coarse-resolution climate models. Further, the strong link between deep convective and stratiform rain and their relationship to large-scale environment variables stresses the need to not isolate deep convection from other rain types when representing their occurrence in climate models. The fact that shallow convection often has an opposite signal compared to the deeper rain types suggests that convective cloud parameterizations need to be adapted to produce both shallow and deep convective rain (currently, GCMs only produce an aggregate convective rain category). 

We used fairly simple spatial covariance functions with common spatial correlation length scale (spatial range) as well as common smoothness for three rain types. We plan to employ more flexible spatial covariance functions, for instance, those that allow latitude dependence of marginal and cross-covariance structures and/or spatially varying smoothness (e.g., \citenp{jun14}).
 
Monte Carlo approximation of log likelihood functions require a large number of simulations of Gaussian random fields, and it naturally enables simple parallelization of the computing. For this work, we used 10 processors that were available for the authors for the parallel computing in R. But with much greater number of processors that will be available soon, we expect much more computationally efficient calculation of approximate likelihood in the near future.

Another future direction of this work is to consider rainfall occurrences as well as the actual rain intensity (or rain rate) data altogether. The actual rain intensity information can be incorporated especially in the covariance modeling of the log-Gaussian intensity function, or rainfall occurrences and rain intensity data altogether can be dealt with under a Marked point process framework.

There may be point process models, other than the LGCP models considered in this paper, that may be suitable for modeling multivariate rainfall occurrences data. For instance, Neyman-Scott clustered point process on spheres used in \cite{lawrence_et_al16} could be extended to multivariate point patterns, and then be compared to the results using the LGCP models. We leave this as one of our future directions of research.  

\newpage

\bibliographystyle{asa}
\bibliography{reference}

\newpage

{\bf \Large Appendix}

\vspace{5mm}

{\bf Proposition.} For $\tilde{\mathbf{S}}$ in \eqref{simulation}, covariance matrix of vec($\tilde{\mathbf{S}}^T$) is a $sN \times sN$ block diagonal matrix. Each block is of size $N \times N$  and equals the approximated version of $\boldsymbol{\Sigma}$ as in \eqref{cov_approxi}. 

\vspace{2mm}

{\it Proof.} Note that elements of $\mathbf{S}_0$ and $\mathbf{S}_1$ are {\it iid} $\mathcal{N}$(0,1). Take any row of $\mathbf{S}_0 \mathbf{B}$ and denote it by $\mathbf{r}_0$ (size $1 \times N$). It is easy to see that $\mathbf{r}_0 = \mathbf{z}_0 \mathbf{B}$, where $\mathbf{z}_0$ is an $1 \times m$ row vector whose elements are {\it iid} $\mathcal{N}$(0,1). Therefore, 
$$\mbox{Var}(\mathbf{r}_0^T) =\mathbf{B}^T~ \mathbf{B} = \mathbf{A} \mathbf{U}_R^{-1} ~(\mathbf{U}_R^{-1})^T \mathbf{A}^T = \mathbf{A} \mathbf{R}^{-1} \mathbf{A}^T.$$     
Now, take any row of $\mathbf{S}_1 \mathbf{U}_V$ and denote it by $\mathbf{r}_1$ (size $1 \times N$). Similarly to the above, 
$$\mbox{Var}(\mathbf{r}_1^T) = \mathbf{U}_V^T~ \mathbf{U}_V = \mathbf{V}.$$
 Therefore, the covariance matrix of a column vector of $\tilde{\mathbf{S}}^T$ equals to $$\mathbf{A} \mathbf{R}^{-1} \mathbf{A}^T + \mathbf{V} =\boldsymbol{W} \approx  \boldsymbol{\Sigma}~~~ \square$$

\end{document}